\documentclass[12pt]{article}
\usepackage{epsfig,amsfonts,amsthm}
\newcommand{\be}{\begin{equation}}
\newcommand{\ee}{\end{equation}}
\newcommand{\bea}{\begin{eqnarray}}
\newcommand{\eea}{\end{eqnarray}}
\newcommand{\dst}{\displaystyle}
\newcommand{\fr}[2]{\frac{{\dst #1}}{{\dst #2}}}

\renewcommand{\Re}{\mathrm{Re }}
\renewcommand{\Im}{\mathrm{Im }}

\newcommand{\lr}[1]{ \langle #1 \rangle}

\def\lsim{\mathrel{\rlap{\lower4pt\hbox{\hskip1pt$\sim$}}
    \raise1pt\hbox{$<$}}}         %less than or approx. symbol
\def\gsim{\mathrel{\rlap{\lower4pt\hbox{\hskip1pt$\sim$}}
    \raise1pt\hbox{$>$}}}         %greater than or approx. symbol

\title{Higgs masses of the general 2HDM in the Minkowski-space formalism}

\author{A.~Deg\'ee$^1$ and I.~P.~Ivanov$^{1,2}$
\\
  {\small $^1$ IFPA, Universit\'{e} de Li\`{e}ge, All\'{e}e du 6 Ao\^{u}t 17, b\^{a}timent B5a, 4000 Li\`{e}ge, Belgium}\\
  {\small $^2$ Sobolev Institute of Mathematics, Koptyug avenue 4, 630090, Novosibirsk, Russia}\\
  }

\begin{document}
\maketitle

\begin{abstract}
We study the masses of the Higgs bosons in the most general two-Higgs-doublet model in a basis-independent approach.
We adapt the recently developed Minkowski-space formalism to this problem and 
calculate traces of any power of the mass-matrix in a compact and reparametrization-invariant form.
Our results can be used to gain insight into the dynamics of the scalar sector of the general 2HDM.
\end{abstract}

\section{Introduction}

\subsection{General 2HDM}

The two-Higgs-doublet model (2HDM) is one of the simplest extensions of the Higgs mechanism of the electroweak symmetry
breaking beyond the Standard-Model, \cite{TDLee,Hunter,CPNSh}. In this model one introduces two
doublets of Higgs fields, $\phi_1$ and $\phi_2$, which interact
with the matter fields and also self-interact via an appropriate Higgs potential.

Higgs potential of 2HDM contains many free parameters, which are not constrained by experiment.
Extensive studies conducted over past decades have shown that playing even with a small subset
of these free parameters one can get a rich spectrum of models with different
phenomenologies (see examples in \cite{CPNSh,ginzreview}).

Recently, it has become clear that not all of these free-parameters are equally important.
One has a certain freedom in choosing the basis in the Higgs field space when writing the lagrangian.
This basis change shifts the values of the parameters of the lagrangian, but by construction it has no effect 
on the physical observables. 
Thus, it is only the {\em basis-invariant features} of the theory, and not the entire
set of free parameters, that really shapes the phenomenology of the model.

When using reparametrization transformations, one is immediately led to the most general 2HDM, 
whose Higgs potential contains all possible electroweak-invariant 
quadratic and quartic combinations of the two doublets.
Several groups have recently focused on the properties of the general 2HDM
and developed a set of basis-invariant tools adequate for this task.
The motivation behind this interest is not to provide the most accurate description of the real world, but rather
to understand the whole spectrum of possibilities offered by the second doublet.
In this way, the most general 2HDM with no {\em a priori} restriction on its free parameters
should be viewed as a useful tool for building specific models with predefined
properties, and it is definitely worth studying in as much detail as possible.

\subsection{Approaches to the general 2HDM}

The main problem with the general 2HDM is that it cannot be worked out with straightforward algebra.
The obstacle arises at the very first step: when minimizing the Higgs potential,
one arrives at algebraic equations of high order, which cannot be solved in the general case.
In this situation, any method that would give any non-trivial insight into the model is welcome.

Following the early suggestion of \cite{CP}, a very elaborate basis-independent
treatment of general 2HDM was presented in \cite{haber} and further developed in \cite{haber2,oneil}.
In this approach one writes the Higgs potential as 
$$
V = Y_{ab} (\phi^\dagger_a \phi_b) + {1\over 2}Z_{abcd}(\phi^\dagger_a \phi_b)(\phi^\dagger_c \phi_d)\,,
$$
and manipulates with $Y_{ab}$ and $Z_{abcd}$ as tensors rather than just a collection of parameters.
Instead of finding explicitly the vector of vacuum expectation values, $v_a$, one adds it
to the set of objects to manipulate with, keeping in mind, however, that it satisfies the extremum condition.
Along these lines, one can find several algebraically independent invariants constructed
as full contractions of the available tensors, and some of the properties of the model could be seen through the prism
of these invariants.

Unfortunately, this powerful technique lacks intuition, as the results arise from lengthy (and often computer-assisted) 
algebra of invariant polynomials. A more appealing approach to the general case was suggested and developed in 
\cite{sartori,nagel,heidelberg,ivanov0,nishi2006,nishi2008}.
In this approach one works not in the space of Higgs fields $\phi_a$, but in the real four-dimensional space of gauge-invariant 
bilinears $(\phi^\dagger_a \phi_b)$ (the orbit space), which has the Minkowski-space signature. 
Many of properties of the Higgs potential can be derived in a very intuitive way based on simple geometric considerations.

This approach was developed further in \cite{ivanov1,ivanov2}, where the reparametrization group
was extended also to non-unitary transformations of the fields. 
In the $1+3$-dimensional orbit space it leads to the full Lorentz group of transformations,
and this freedom provides even more insight into the properties of the 2HDM potential.
In particular, many of the statements about the general 2HDM are much more naturally formulated 
in terms of the four eigenvalues of the Minkowski tensor $\Lambda^{\mu\nu}$ (see the next Section) rather than its
space-like part.
Another key point of \cite{ivanov1,ivanov2} was that all essential results were formulated
exclusively in terms of the parameters of the potential, without using the yet-unknown
vacuum expectation values.
For example, within this approach one could formulate conditions for the spontaneous $CP$-violation
and draw the full phase diagram of the model
exclusively in term of the parameters of the potential, without using the unknown vacuum expectation values.

Thanks to all these approaches, we have now a fairly detailed understanding of the properties
of the Higgs potential and of the vacuum in the general 2HDM. 

\subsection{Towards dynamics of the general 2HDM}

The next step in the study of the general 2HDM is to understand its dynamics.
This includes the mass spectrum of the physical Higgs bosons, the pattern of their interactions,
as well as their couplings to the fermions. All this must be done within a basis-independent approach.

Let us stress once again that if one chooses a restricted Higgs potential, for example,
an explicitly $CP$-symmetric one, the entire calculation is drastically simplified. 
One can explicitly find the minimum of the potential and calculate the masses and the interaction of the Higgs bosons.
This straightforward approach fails for the most general 2HDM, which calls upon more involved techniques 
for the analysis of its properties.

The mass spectrum of the general 2HDM was studied in a number of recent papers.
For example, in \cite{heidelberg,nishi2008,ivanov1} the mass matrix was explicitly calculated
in a specific basis and not in a reparametrization-invariant form.
An interesting study was also presented in \cite{ferreira-masses}, where certain
bounds and relations between the masses and the parameters of the potential were observed,
however that work relied only on numerical analysis.
Finally, very recently a very detailed account of the dynamics of the general 2HDM 
was presented in \cite{oneil}. Among other results, explicit expressions of the mass matrix 
were derived in $U(2)$-invariant way in terms of various full contractions 
of tensors $Y_{ab}$ and $Z_{abcd}$ as well as vacuum expectation values.

In the present paper we show how to analyze the masses of the physical Higgs bosons 
in the Minkowski-space formalism.
We obtain compact $SO(1,3)$-invariant expressions for the traces of any power of the mass matrix.
Thus, we can now use the full power of the extended reparametrization symmetry of the problem 
to gain further insight into dynamical properties of the general 2HDM.\\

The structure of the paper is the following. In Section 2 we briefly review the 
Minkowski-space approach to the general 2HDM.
The main results of the paper are then derived in Section 3. 
For each possible type of the 2HDM vacuum, we find the mass-matrix
and calculate the traces of its powers in reparametrization-covariant way.
A discussion of the results and conclusions are presented in Section 4.

\section{Overview of the formalism}
In this work we focus on the scalar sector of 2HDM.
The Higgs potential of the most general renormalizable 2HDM, $V_H = V_2 + V_4$, is
conventionally parametrized as
\bea
V_2&=&-{1\over 2}\left[m_{11}^2(\phi_1^\dagger\phi_1) +
m_{22}^2(\phi_2^\dagger\phi_2)
+ m_{12}^2 (\phi_1^\dagger\phi_2) + m_{12}^{2\ *} (\phi_2^\dagger\phi_1)\right]\,;\nonumber\\
V_4&=&\fr{\lambda_1}{2}(\phi_1^\dagger\phi_1)^2
+\fr{\lambda_2}{2}(\phi_2^\dagger\phi_2)^2
+\lambda_3(\phi_1^\dagger\phi_1) (\phi_2^\dagger\phi_2)
+\lambda_4(\phi_1^\dagger\phi_2) (\phi_2^\dagger\phi_1) \label{potential}\\
&+&\fr{1}{2}\left[\lambda_5(\phi_1^\dagger\phi_2)^2+
\lambda_5^*(\phi_2^\dagger\phi_1)^2\right]
+\left\{\left[\lambda_6(\phi_1^\dagger\phi_1)+\lambda_7
(\phi_2^\dagger\phi_2)\right](\phi_1^\dagger\phi_2) +{\rm
h.c.}\right\}\,.\nonumber
\eea
It contains 14 free parameters,
four in the quadratic and 10 in the quartic terms,
which makes the phenomenology of 2HDM very rich even at tree level.
However, not all points in this 14-dimensional space of parameters lead to distinct physics:
if two sets of parameters can be mapped into each other by a certain
linear transformation between the doublets (reparametrization transformation, or basis change), 
they will lead to the same physics, \cite{haber,ginzreview}.
Usually one insists that the kinetic term be invariant, so one considers 
only global unitary transformations between the two doublets, $U(2)$.
However, as shown in \cite{ivanov1,ivanov2}, one can extend this reparametrization group
to the general linear group $GL(2,C)$. The Higgs kinetic terms is not invariant under
non-unitary transformations, but it can be treated in a reparametrization-covariant way, 
so that all the physical observables still remain invariant under this extended reparametrization group.
This approach has provided several new insights, which would be very difficult to see
using the more traditional unitary reparametrization group.

Technically, the extended reparametrization group can be implemented as follows.
We switch from the fields to bilinears and introduce the four-vector
$r^\mu = (r_0,\,r_i) = (\Phi^\dagger \Phi,\, \Phi^\dagger \sigma^i \Phi)$,
where $\Phi = (\phi_1,\,\phi_2)^T$ is a 2-dimensional vector of Higgs doublets
and $\sigma^i$ are the Pauli matrices. This four-vector is gauge invariant
and parametrizes the gauge orbits in the space of the Higgs fields.
The general reparametrization group $GL(2,C)$ can be written as $\mathbb{C}^*\otimes SL(2,C)$,
where $\mathbb{C}^*$ is the group of simultaneous multiplication of both $\phi_i$ with the same complex number,
while $SL(2,C)$ is the special linear transformation group. 
It is the latter group that leads to non-trivial transformations of the Higgs potential,
which we now focus on.

Transformations of $\Phi$ under $SL(2,C)$ correspond to the $SO(1,3)$ transformations of $r^\mu$, 
equipping the gauge orbit space with the Minkowski-space structure.
It follows from the definition of $r^\mu$ that
\be
r_0 = (\phi_1^\dagger\phi_1)+ (\phi_2^\dagger\phi_2) \ge 0\,,\label{from-definition}\quad
r^\mu r_\mu = 4\left[(\phi_1^\dagger\phi_1)(\phi_2^\dagger\phi_2)- (\phi_1^\dagger\phi_2)(\phi_2^\dagger\phi_1)\right]\ge 0\,, 
\ee
so that the physically realizable vectors $r^\mu$
populate not the entire $1+3$-dimensional Minkowski space, but the future lightcone ($LC^+$).
The Higgs potential in the $r^\mu$-space can be written in a very compact form:
\be
V = - M_\mu r^\mu + {1\over 2}\Lambda_{\mu\nu} r^\mu r^\nu\,.\label{Vmunu}
\ee
Here the four-vector $M_\mu$ is built from parameters $m_{ij}^2$ in (\ref{potential}),
while the symmetric four-tensor $\Lambda_{\mu\nu}$ is constructed from
the quartic coefficients $\lambda_i$.
Their explicit expressions as well as some properties can be found in \cite{nishi2006,ivanov1,ivanov2}.
Here we just note the most important property of $\Lambda_{\mu\nu}$ for potentials stable in a strong 
sense\footnote{We use here the terminology of \cite{heidelberg}: the potential is stable in a strong sense,
if its quartic part increases along all rays starting from the origin in the Higgs field space.
The potential is called stable in a weeak sense, if the quartic part has flat directions, but the quadratic potential
increases along them. For the Minkowski-space analysis of potentials stable in a weak sense, see \cite{GL},
where a similar condensed-matter problem was considered.}:
$\Lambda_{\mu\nu}$ can always be diagonalized by a certain $SO(1,3)$ transformation of the $r^\mu$-space,
and after diagonalization it takes form
\be
\Lambda_{\mu\nu} = \mathrm{diag}(\Lambda_0,\,-\Lambda_1,\,-\Lambda_2,\,-\Lambda_3)\quad
\mathrm{with} \quad \Lambda_0>0,\quad \Lambda_0>\Lambda_i\,,\ i=1,2,3\,,\label{Lambdamunudiag}
\ee
where the inequalities among the eigenvalues result from the positivity constraint on the potential.
The minus signs in front of the ``space-like'' eigenvalues arise from the pseudo-euclidean
metric in the orbits space.

It is known that the potential (\ref{Vmunu}) can have three types of minima:
(i) the electroweak (EW) conserving,
(ii) the EW-breaking but charge conserving (i.e. neutral), and
(iii) the EW- and charge-breaking ones.
One can use the v.e.v.s of the two doublets $\lr{\phi_i}$ to construct $\lr{r^\mu}$.
Then, the three type of minima correspond to:
(i) $\lr{r^\mu} = 0$ (the apex of the forward lightcone $LC^+$),
(ii) $\lr{r^\mu} \not = 0$ but $\lr{r^\mu}\lr{r_\mu} = 0$ (the surface of $LC^+$),
(iii) $\lr{r^\mu} \not = 0$ and $\lr{r^\mu}\lr{r_\mu} > 0$ (the interior of $LC^+$).
The position of the charge-breaking extremum $\lr{r_\nu}_{ch}$ is given by the following
equations:
\be
\Lambda^{\mu\nu} \lr{r_\nu}_{ch} = M^\mu\,,\label{extremum1}
\ee
If $\Lambda^{\mu\nu}$ is not singular, a solution of this system always exists and is unique: 
$\lr{r_\mu}_{ch} = m_\mu \equiv (\Lambda^{-1})_{\mu\nu}M^\nu$.
However, the requirement that $\lr{r_\nu}_{ch}$ lies inside the forward lightcone
places bounds on $M^\mu$ that could yield physically realizable solutions.
In addition, the charge-breaking extremum is minimum only if all $\Lambda_i < 0$, $i=1,2,3$,
i.e. if the tensor $\Lambda_{\mu\nu}$ is positive-definite in the entire space of non-zero vectors $r^\mu$.
When searching for the neutral extrema, we use the Lagrange multiplier technique.
The positions of all neutral extrema $\lr{r^\mu}$ are the solutions of the following simultaneous equations:
\be
\Lambda^{\mu\nu} \lr{r_\nu} - \zeta^\mu = M^\mu\,,\quad \zeta^\mu \equiv \zeta \lr{r^\mu}\,,\label{extremum2}
\ee
where $\zeta$ is a Lagrange multiplier. This system can have up to six solutions, \cite{nagel,heidelberg,ivanov1},
among which there are at most two minima, while the other are saddle points, \cite{ivanov2}.

Finally, following \cite{ivanov1}, we write the Higgs kinetic term covariantly as
\be
K = K_\mu \rho^\mu\,,\quad \rho^\mu = (\partial_\alpha \Phi)^\dagger \sigma^\mu (\partial^\alpha \Phi)\,,
\ee
where $\alpha$ denotes the usual space-time coordinates, while $\mu$, as before, refers to the orbit space.
The reparametrization transformation properties of $\rho^\mu$ are the same as for $r^\mu$.
In the ``default'' frame, $K^\mu = (1,\,0,\,0,\,0)$. Upon an $SO(1,3)$ transformation, $K^\mu$
acquires non-zero ``space-like'' coordinates, however the condition $K^\mu K_\mu = 1$ is always satisfied.
The four-vector $K^\mu$ is not involved in the search for the minimum of the potential,
however it affects the mass matrix at this minimum.
This generalized kinetic term effectively incorporates the non-diagonal kinetic term, which, as was
argued in \cite{ginz-kinetic}, must be introduced in the initial lagrangian to restore renormalizability of the model.

\section{Mass matrix of the most general 2HDM} 

In the previous studies, \cite{ivanov1,ivanov2} the Minkowski-space formalism
was used to understand various properties of the 2HDM lagrangian and
of the vacuum state. The next logical step is to study the dynamics of the model in an reparametrization-covariant way.
In this paper we make a step towards fulfilling this program. We obtain expressions for 
the mass matrices of the physical Higgs bosons in the most general 2HDM and study some of their properties. 

When doing so, we stick to the Minkowski space formalism, but we adapt it to our problem.
Although the masses are physical observables and are reparametrization-invariant,
the mass-matrix is, obviously, basis-dependent.  
So, for intermediate calculations we switch back from the bilinears to the Higgs fields themselves,
derive the mass-matrix in a specific basis, then calculate the traces of the powers of this matrix,
and return to the Minkowski-space formalism. 
Although the resulting equations do not yield the masses in a simple closed form, 
they nevertheless can be useful for the analysis of the general 2HDM.

In the subsection devoted to the neutral vacuum below, we also comment 
on relation of our results with some of the previous studies of the mass spectrum.

\subsection{Switching to the real fields}

Let us denote the complex fields as $\phi_{i,\alpha}$, where $i=1,2$ indicates the doublet, 
while $\alpha=\uparrow,\downarrow$ indicates the upper and lower components in a given doublet.
Let us then introduce the 8-component real vector of scalar fields $\varphi_a$, $a=1, ..., 8$, with the following
components:
\be
\varphi_a = \left(
\Re\phi_{1,\uparrow},\, \Im\phi_{1,\uparrow},\, \Re\phi_{2,\uparrow},\, \Im\phi_{2,\uparrow},\, 
\Re\phi_{1,\downarrow},\, \Im\phi_{1,\downarrow},\, \Re\phi_{2,\downarrow},\, \Im\phi_{2,\downarrow}
\right)\,.
\ee
The four-vector $r^\mu$ can be rewritten in terms of $\varphi_a$ as
\be
r^\mu = \varphi_a \Sigma^\mu_{ab} \varphi_b\,.\label{definitionSigma}
\ee
Here, $\Sigma^\mu$ are four real symmetric 8-by-8 matrices; $\Sigma^0$ is just the unit matrix,
while explicit form of $\Sigma^i$ can be immediately reconstructed from the definitions (see Appendix).
Since the upper and lower components of the doublets are not mixed by the Higgs potential, matrices $\Sigma^\mu$ have
a block-diagonal form, composed of identical 4-by-4 matrices. Below, we will often deal with these 4-by-4 matrices,
denoting them by the same letter $\Sigma^\mu$. Which set of matrices is being used, 4-by-4 or 8-by-8, should be clear from the context.

In contrast to $\sigma^\mu$, the matrices $\Sigma^\mu$ do not form a closed algebra,
but they belong to a larger algebra $(\Sigma^\mu,\Pi^\mu)$, described in the Appendix.
They also share with $\sigma^\mu$ an important property:
\be
\left\{\Sigma^i,\Sigma^j\right\} = 2 \delta^{ij}\cdot \mathbb{I}_8\,,
\ee
where brackets denote the anticommutator.
It follows then that if a regular real symmetric 8-by-8 matrix $A$ is written as $a_\mu \Sigma^\mu$,
then its inverse is
\be
A^{-1} = { a_\mu \bar\Sigma^\mu \over a_\mu a^\mu}\,, \quad \bar\Sigma^\mu \equiv (\Sigma^0,\, -\Sigma^i)\,.
\ee
Below we will encounter products of matrices $\Sigma$'s and $\bar\Sigma$'s. When simplifying these products,
the following results prove useful:
\bea
&& {1 \over 2}(\Sigma^\mu\bar\Sigma^\nu+\bar\Sigma^\nu \Sigma^\mu) = g^{\mu\nu}\cdot \mathbb{I}_8\,,\label{propertiesSigma}\\
&& {1 \over 2}\left(\Sigma^\mu \bar\Sigma^\rho \Sigma^\nu + \Sigma^\nu \bar\Sigma^\rho \Sigma^\mu\right)
= g^{\mu\rho}\Sigma^\nu + g^{\nu\rho}\Sigma^\mu - g^{\mu\nu}\Sigma^\rho\,.\nonumber
\eea
With this notation, we can give a compact expression for the mass matrix in a specific basis.
Let us write the expansion of the scalar lagrangian near an extremum as
$$
L \approx (K_\rho \Sigma^\rho_{ab})(\partial_\alpha\varphi_a)(\partial^\alpha\varphi_b) 
- H_{ab}(\varphi_a-\lr{\varphi_a})(\varphi_b-\lr{\varphi_b})\,,
\quad H_{ab} \equiv {1\over 2} {\partial^2 V \over \partial\varphi_a\,\partial\varphi_b}\,,
$$
where the hessian $H_{ab}$ is calculated at the extremum.
The 8-by-8 mass matrix can then be expressed as
\be
{\cal M}_{ac} = (K_\rho \Sigma^\rho)_{ab}^{-1} H_{bc} = K_\rho \bar\Sigma^\rho_{ab} H_{bc}\,.\label{massmatrix}
\ee
In the rest of this Section we calculate this mass matrix and analyze its eigenvalues for the three possible types
of vacua: electroweak-symmetric, charge-breaking and neutral.

\subsection{Electroweak-symmetric vacuum}

The masses of the Higgs bosons in the electroweak-symmetric vacuum are determined only by the
quadratic term of the potential and can be easily calculated in a straightforward way. 
The eight masses are grouped into two quartets with values
(\ref{potential}) are
\be
m_{1,2}^2 = {1 \over 4}\left((-m_{11}^2)+(-m_{22}^2) \pm \sqrt{(m_{11}^2-m_{22}^2)^2+4|m_{12}|^2}\right)\,.\label{massEWsimple}
\ee
These masses squared are positive, if $m_{11}^2 < 0$, $m_{22}^2 < 0$ and $m_{11}^2m_{22}^2 > |m_{12}|^2$.
However, we find it useful to work out this simple case in the reparametrization-covariant formalism just to 
illustrate how it works.

The hessian $H_{ab}$ comes only from the $M_\mu r^\mu$ term of the potential and is equal to $-M_\mu \Sigma^\mu_{ab}$.
The mass matrix is then
\be
{\cal M}_{ab} = K_\rho (-M_\mu) (\bar\Sigma^\rho\Sigma^\mu)_{ab}\,.\label{massEW1}
\ee
Matrices $\Sigma$'s have a block-diagonal form, and therefore so does the mass matrix (\ref{massEW1}). It is built
of two identical 4-by-4 blocks $({\cal M}_4)_{ab}$, with $a,b=1,2,3,4$, whose form is still given by the same 
expression but now with 4-by-4 matrices $\Sigma^\mu$. In order to find its eigenvalues, let us calculate
the trace of its successive powers:
\bea
\mathrm{Tr}[{\cal M}_4] &=& K_\rho (-M_\mu) \mathrm{Tr}[\bar\Sigma^\rho\Sigma^\mu] = -4(KM)\,,\nonumber\\
\mathrm{Tr}[({\cal M}_4)^2] &=& K_\rho (-M_\mu) K_{\rho'} (-M_{\mu'}) \mathrm{Tr}[\bar\Sigma^\rho\Sigma^\mu\bar\Sigma^{\rho'}\Sigma^{\mu'}]
\nonumber\\
&=& 2(KM) K_\rho M_\mu \mathrm{Tr}[\bar\Sigma^\rho\Sigma^\mu] - K_\rho K_{\rho'} M^2 \mathrm{Tr}[\bar\Sigma^\rho\Sigma^{\rho'}]
= 8 (KM)^2 - 4 K^2 M^2\,,\nonumber\\
\mathrm{Tr}[({\cal M}_4)^n] &=& -2(KM)\mathrm{Tr}[({\cal M}_4)^{n-1}] - K^2 M^2 \mathrm{Tr}[({\cal M}_4)^{n-2}]\,.  
\eea
These relations among the traces prove the mass matrix has only two independent eigenvalues. 
A simple analysis shows that there are two pairs
of different eigenvalues, which are equal to
\be
m_{1,2}^2 = -(KM) \pm \sqrt{(KM)^2 - M^2}\,,
\ee
where we used $K^2=1$.
This expression is reparametrization-invariant and can be calculated in any frame. 
In particular, in the original frame, where $K^\mu = (1,0,0,0)$,
we obtain
\be
m_{1,2}^2 = -M_0 \pm |\vec M|\,.\label{massEW3}
\ee
Using the definition of $M^\mu$, one can immediately recover (\ref{massEWsimple}).
Eq.~(\ref{massEW3}) also shows that in order for the EW-symmetric extremum to be minimum,
the four-vector $M^\mu$ must lies inside the backward lightcone.

\subsection{Charge-breaking vacuum}

Let us now find the mass matrix of the general 2HDM in the case of a charge-breaking vacuum.
The hessian has the following form:
\be
H_{bc} = 2\Lambda_{\mu\nu}\Sigma^\mu_{bb'} \Sigma^\nu_{cc'} \varphi_{b'}\varphi_{c'}\,.\label{hessian-ch}
\ee
All fields here must be understood as v.e.v.'s $\lr{\varphi_a}$, but to keep the notation simple, we will suppress
the brackets. Thus, the mass matrix can be written as
\be
{\cal M}_8 = 2K_\rho \Lambda_{\mu\nu} \bar\Sigma^\rho \Sigma^\mu (\varphi\otimes\varphi) \Sigma^\nu\,.\label{mass-ch}
\ee
By construction, this is a 8-by-8 matrix. However, we know that it must have four flat directions corresponding
to the Goldstone modes. We shall now get rid of these four flat directions by showing that 
there exists a 4-by-4 matrix ${\cal M}_4$ such that trace of any power of ${\cal M}_8$ is equal
to the trace of the same power of ${\cal M}_4$. 

Indeed, consider the trace of ${\cal M}_8$. Thanks to the properties of $\Sigma$'s, we have
\bea
\mathrm{Tr}\left[{\cal M}_8\right]
&=& 2K_\rho \Lambda_{\mu\nu}\, \varphi \Sigma^\nu \bar\Sigma^\rho \Sigma^\mu \varphi
= 2K_\rho \Lambda_{\mu\nu} \left(g^{\mu\rho} m^\nu + g^{\nu\rho} m^\mu - g^{\mu\nu}m^\rho\right) \nonumber\\
&=& 2K_\rho m_\mu \left(2 \Lambda_{\rho\mu} - \mathrm{Tr}\Lambda\, g_{\rho\mu}\right)\equiv 2 \mathrm{Tr}\left[S \cdot \Lambda\right]\,.\nonumber
\eea
Here, the matrix $S \cdot \Lambda$ is a symbolic form of the tensor $S^{\mu}{}_\alpha\Lambda_{\alpha}{}_\nu\equiv S^{\mu\alpha}\Lambda_{\alpha\nu}$, where 
\be
S^{\nu\mu} \equiv K^\nu m^\mu + K^\mu m^\nu - (Km) g^{\nu\mu}\,. \label{Smunu}
\ee
Note that the matrix $S \cdot \Lambda$ is defined in the euclidean space, 
and although it contains the tensors $S^{\mu\alpha}$ and $\Lambda_{\alpha\nu}$, they are contracted 
according to the usual rules of matrix multiplication. 

Consider now the trace of the square of ${\cal M}_8$:
\be
\mathrm{Tr}\left[({\cal M}_8)^2\right] =
4K_\rho \Lambda_{\mu\nu} K_{\rho'} \Lambda_{\mu'\nu'}\cdot  
\varphi \Sigma^\nu \bar\Sigma^{\rho'} \Sigma^{\mu'} \varphi \cdot \varphi \Sigma^{\nu'} \bar\Sigma^\rho \Sigma^\mu \varphi\,.
\ee
Note that this expression does not factorize because $\Lambda_{\mu\nu}$ and $\Lambda_{\mu'\nu'}$ couple
the first and the second threads of $\Sigma$'s.
Consider one of these threads, e.g.
$$
\varphi_a(\Sigma^\nu \bar\Sigma^{\rho'} \Sigma^{\mu'})_{ab} \varphi_b\,.
$$
This is a quadratic form in $\varphi_a$; therefore, only the $ab$-symmetric part of the product of $\Sigma$'s
survives. This effectively leads to the $\nu \leftrightarrow \mu'$ symmetrization, and one can again apply (\ref{propertiesSigma})
to obtain
$$
K_{\rho'} \varphi_a \left(\Sigma^\nu \bar\Sigma^{\rho'} \Sigma^{\mu'}\right)_{ab} \varphi_b = S^{\nu\mu'}\,.
$$
The trace of the square of the mass matrix is then
$$
\mathrm{Tr}\left[({\cal M}_8)^2\right] = 4 \Lambda_{\mu\nu}S^{\nu\mu'} \Lambda_{\mu'\nu'}S^{\nu'\mu} =
\mathrm{Tr}\left[(2 S\cdot\Lambda)^2\right]\,.
$$
This calculation is easily generalizes to any power of the mass matrix:
\be
\mathrm{Tr}\left[({\cal M}_8)^n\right] = \mathrm{Tr}\left[(2 S\cdot\Lambda)^n\right]\,.
\ee
The fact that the trace of any power of ${\cal M}_8$ is equal to the trace of the same power of the 4-by-4 matrix
$2S \cdot \Lambda$, means that there are four zero-modes in ${\cal M}_8$ and that all the four non-zero eigenvalues of ${\cal M}_8$ 
coincide with the eigenvalues of $2 S \cdot \Lambda$.
Thus, the four eigenvalues of the matrix $2 S \cdot \Lambda$ gives the masses squared of the physical Higgs bosons in the charge-breaking
vacuum.

There is no simple way to calculate the masses themselves. However, the product of all four masses squared
can be easily inferred from the above expression:
\be
\prod_{i} m_i^2 = \mathrm{det}(2S\cdot\Lambda) = 16\, \mathrm{det}S\cdot \mathrm{det}\Lambda\,.
\ee
Both tensors here are written in the euclidean space.
Determinant of euclidean ${\Lambda^\alpha}_\beta$ is the product of the eigenvalues\footnote{Note a subtlety here: 
in a generic basis, the eigenvalues of the euclidean matrix ${\Lambda^\alpha}_\beta$, which is not even symmetric, 
are different from the eigenvalues of the Minkowski tensor $\Lambda^{\mu\nu}$, 
i.e. $\Lambda_0$ and $\Lambda_i$. However, the product of all the eigenvalues of these two matrices are equal.} 
of Minkowski $\Lambda^{\mu\nu}$:
det$\Lambda = \Lambda_0\Lambda_1\Lambda_2\Lambda_3$.
In order to calculate the other determinant, let us take a closer look at $S^{\mu\nu}$.
The way it is defined, Eq.~(\ref{Smunu}), allows us to immediately find its eigenvalues.
Indeed, consider first a reduced version of this tensor, $K^\mu m^\nu+K^\nu m^\mu$.
In general, $K^\mu$ and $m^\mu$ are non-parallel four vectors, both lying strictly inside the forward lightcone.
Within the subspace spanned by them, one can identify two eigenvectors of this reduced tensor,
$$
e^\mu_{\pm} = {K^\mu \over \sqrt{K^2}} \pm {m^\mu \over \sqrt{m^2}}\,, \quad e^\mu_{+} e_{-\,\mu} = 0\,,
$$
whose eigenvalues are $(Km) \pm \sqrt{K^2 m^2}$. Note that $e_+^\mu$ lies inside the forward lightcone,
while $e_-^\mu$ lies outside it.
In addition, there are two eigenvectors in the subspace orthogonal to $K^\mu$ and $m^\mu$, with zero eigenvalues.
Since adding a term proportional to $g_{\mu\nu}$ does not change the eigenvectors but just
shifts all the eigenvalues by a common constant,
we get the following result: $S^{\mu\nu}$ is diagonalizable by an appropriate $SO(1,3)$ transformation,
and after diagonalization it take form:
\be
S^{\mu\nu}=\mathrm{diag}(S_0,\,-S_1,\,-S_2,\,-S_3)\,,\quad
S_0 = \sqrt{m^2}\,,\ S_1 = -\sqrt{m^2}\,, \ S_2 = S_3 = - (Km)\,.\label{Seigenvalues}
\ee
Therefore, we obtain:
\be
\prod_{i} m_i^2 = 16 \Lambda_0(-\Lambda_1)(-\Lambda_2)(-\Lambda_3) \cdot m^2 (Km)^2\,.\label{product}
\ee
As said above, a charge-breaking extremum exists, if $m^\mu$ lies inside the future lightcone, i.e.
if $m^2>0$ and $(Km)>0$.
It is also known that the charge-breaking extremum is a minimum if
the tensor $\Lambda^{\mu\nu}$ is positive-definite in the entire Minkowski space,
i.e. if all its spacelike eigenvalues $\Lambda_{1,2,3}$ are 
negative\footnote{We checked that these conditions can be also inferred from the positive-definiteness of the mass matrix
just derived.}.
Thus, all factors in (\ref{product}) are positive.

Another observation concerns cases when the potential has an explicit symmetry.
Consider, for example, the lowest possible explicit symmetry, a $Z_2$-symmetry\footnote{This symmetry
is known in the literature as a generalized $CP$-symmetry. The ``conventional $Z_2$'' corresponds, strictly speaking,
to a $(Z_2)^2$-symmetry of the potential, see details in \cite{ivanov2}.}, which consists in reflection
of, say, second axis.
This explicit symmetry means that $K_2=0$, $M_2 =0$, and that $\Lambda_{2\mu}=0$ for $\mu \not = 2$.
It is known that the position of the charge-breaking minimum preserves all the discrete symmetries, so that
$m_2$ is also zero. 
In this case one can immediately calculate the mass squared of the excitation that violates this symmetry: 
\be
m_2^2 = 2(-\Lambda_2)(Km)\,.
\ee

\subsection{Neutral vacuum}
Let us now consider the neutral vacuum. The four-vector $r^\mu$ corresponding
to a neutral vacuum must lie on the surface of the forward lightcone 
(again, we always refer to the v.e.v.'s, so that the brackets $\lr{\cdots}$ are implicitly assumed). 
Therefore, the minimization procedure involves a Lagrange multiplier $\zeta$,
which brings up a new lightcone four-vector, $\zeta_\mu$, defined as $\zeta_\mu = \Lambda_{\mu\nu}r^\nu - M_\mu = \zeta\cdot r_\mu$.
This new four-vector gives rise to an additional term in the mass matrix:
\be
{\cal M}_8 = 2K_\rho \Lambda_{\mu\nu} \bar\Sigma^\rho \Sigma^\mu (\varphi\otimes\varphi) \Sigma^\nu +
K_\rho \zeta_\mu \bar\Sigma^\rho \Sigma^\mu\,.\label{mass-n}
\ee
This matrix is again an 8-by-8 real symmetric matrix. However, one can easily split it into
two 4-by-4 matrices corresponding to the charged (the first four components of $\varphi_a$) 
and neutral (the last four components of $\varphi_a$) modes, which do not mix.

Before we proceed, let us note that essentially this expression for the mass matrix of the most
general 2HDM, but with a trivial kinetic part,
was obtained in other works, \cite{heidelberg,ivanov1,oneil}. All these papers followed then 
the standard procedure: one switches to the basis where only the first doublet
has non-zero v.e.v. (the Higgs basis), and then the entries of the mass matrix can then be written in a simple
way via the parameters of the potential in this specific basis as well as $v^2$.
We show in this subsection that the basis-invariant features of the mass matrix
can be written in an $SO(1,3)$-covariant way, without referring to any specific basis.
The power of the covariant expression is that it can be analyzed in any desired basis,
e.g. in the $\Lambda_{\mu\nu}$-diagonal basis.
We checked that in the canonical basis, our result reproduce those of \cite{ivanov1,oneil}.

Consider first the charged excitations. Their masses arise solely from the last term in (\ref{mass-n}):
\be
{\cal M}^{ch.}_4 = K_\rho \zeta_\mu \bar\Sigma^\rho \Sigma^\mu\,,
\ee
where $\Sigma$'s are now 4-by-4 matrices. By explicit calculations and using the fact that $\zeta^2 = 0$,
we get:
\be
\mathrm{Tr}{\cal M}^{ch.}_4 = 4(K\zeta)\,,\quad
\mathrm{Tr}[({\cal M}^{ch.}_4)^2] = 8(K\zeta)^2\,,\quad
\mathrm{Tr}[({\cal M}^{ch.}_4)^n] = 2[2(K\zeta)]^n\,.
\ee
It means that this matrix has only two non-zero eigenvalues, which are identical and equal to
\be
m_{H^\pm}^2 = 2(K\zeta)\,.
\ee
This implies, in particular, that in order for the extremum to be minimum, $\zeta$ must lie on the surface of the forward, 
not backward lightcone.

For the neutral modes one has the same expression as in (\ref{mass-n}), but with 4-by-4 matrices $\Sigma^\mu$:
\be
{\cal M}^{n.}_4 = 2K_\rho \Lambda_{\mu\nu} \bar\Sigma^\rho \Sigma^\mu (\varphi\otimes\varphi) \Sigma^\nu +
K_\rho \zeta_\mu \bar\Sigma^\rho \Sigma^\mu\,.\label{mass-n-n}
\ee
Let us calculate the trace of the mass matrix of the neutral Higgs bosons:
\bea
\mathrm{Tr}{\cal M}^{n.}_4 &=& 2 K_\rho \Lambda_{\mu\nu} \varphi \Sigma^\nu \bar\Sigma^{\rho} \Sigma^{\mu} \varphi + 4 (K\zeta) =
4\Lambda_{\mu\nu}K^\mu r^\nu - 2 \mathrm{Tr}\Lambda\, (K r) + 4 (K\zeta)\nonumber\\
&&=2(4\Lambda_{\mu\nu}-\mathrm{Tr}\Lambda\,g_{\mu\nu})K^\mu r^\nu - 4 (KM)\,.\label{TrMn4}
\eea
We expect that among the four neutral modes there will be one goldstone, which makes the determinant
of ${\cal M}^{n.}_4$ zero. To check it explicitly, we first factor out the matrix $K_\rho \bar\Sigma^\rho$
and check by a direct calculation that its determinant is equal to $(K_\mu K^\mu)^2 = 1$. The remaining
determinant
$$
\mathrm{det}\left[2\Lambda_{\mu\nu} \Sigma^\mu (\varphi\otimes\varphi) \Sigma^\nu + \zeta_\mu \Sigma^\mu\right]
$$
is equal to zero, which can be best seen in the Higgs basis, where the second row and the second column have only zeros.
In the generic basis, the goldstone mode is $w_i = (\Pi^0)_{ij} \phi_j$, where the matrix $\Pi^0$ is the generator 
of the $SO(2)$ rotations between the real and imaginary parts, see Appendix.

\subsection{The extra symmetry of the neutral modes}

The appearance of the tensor $4\Lambda_{\mu\nu}-\mathrm{Tr}\Lambda\,g_{\mu\nu}$ in (\ref{TrMn4}) is not accidental,
but reflects an extra symmetry of the neutral mass matrix.
If we consider the neutral vacuum and if we analyze only neutral excitations, we always stay on the surface of the lightcone:
we consider only $r^\mu$ such that $g_{\mu\nu}r^\mu r^\nu = 0$.
This means that if we shift the tensor $\Lambda_{\mu\nu}$ in the potential as
\be
\Lambda_{\mu\nu} \to \Lambda_{\mu\nu} + C g_{\mu\nu}\label{extrasymmetry}
\ee
with an arbitrary $C$, the purely neutral contribution to the potential does not change,
and neither does the neutral mass matrix. The tensor $4\Lambda_{\mu\nu}-\mathrm{Tr}\Lambda\,g_{\mu\nu}$
is precisely the combination that is invariant under such a shift. In terms of the original parametrization of the quartic
potential (\ref{potential}), this symmetry means that the neutral Higgs boson masses do not depend on
the value of Tr$\Lambda = \lambda_3-\lambda_4$.

One can make use of this extra symmetry to simplify the neutral Higgs boson mass matrix.
First, note that the neutral mass matrix (\ref{mass-n-n}) is invariant under the transformation (\ref{extrasymmetry}) thanks to
the following relation:
\be
2 g_{\mu\nu} \Sigma^\mu (\varphi\otimes\varphi) \Sigma^\nu + r_\mu \Sigma^\mu = 0\,.
\ee
Let us recall now that $\zeta_\mu$ is proportional to $r_\mu$: $\zeta_\mu = \zeta\cdot r_\mu$,
where $\zeta$ is the Lagrange multiplier of the minimization problem.
Then, we can group the two terms in (\ref{mass-n-n}) together:
\be
{\cal M}^{n.}_4 = 2K_\rho {\tilde\Lambda}_{\mu\nu} \bar\Sigma^\rho \Sigma^\mu (\varphi\otimes\varphi) \Sigma^\nu
\quad\mbox{where}\quad {\tilde\Lambda}_{\mu\nu} \equiv \Lambda_{\mu\nu}-\zeta g_{\mu\nu}\,.\label{mass-n-n2}
\ee
It is remarkable that the new tensor ${\tilde\Lambda}_{\mu\nu}$ is itself invariant under (\ref{extrasymmetry})
as this shift is accompanied by $\zeta \to \zeta+C$:
\be
\zeta r^\mu \equiv \zeta^\mu = \Lambda^{\mu\nu} r_\nu - M_\mu \to (\Lambda^{\mu\nu} +C g^{\mu\nu}) r_\nu - M_\mu
= \zeta^\mu + C r^\mu = (\zeta+C)r^\mu\,.
\ee
With this expression in hand, we can again use the trick from the analysis of the charge-breaking vacuum 
and state that all the neutral boson masses 
are given by the eigenvalues of the following matrix written in a manifestly covariant form:
\be
{\tilde {\cal M}}^{n.}_4 = 2 {\tilde S}\cdot {\tilde \Lambda}\,,\quad\mbox{where}\quad 
{\tilde S}^\mu{}_\nu = K^\mu r_\nu + K_\nu r^\mu - (Kr)\delta^{\mu}_{\nu}\,.
\ee
Therefore, one can immediately write the trace of any power of the mass matrix:
\be
\mathrm{Tr}[({\cal M}^{n.}_4)^k] = 2^k {\tilde S}^{\mu_1}{}_{\nu_1}{\tilde\Lambda}^{\nu_1}{}_{\mu_2}\cdots
{\tilde S}^{\mu_k}{}_{\nu_k}{\tilde\Lambda}^{\nu_k}{}_{\mu_1}\,,
\ee
and calculate the determinant of ${\tilde S}^\mu{}_{\nu}$ using (\ref{Seigenvalues}):
\be
\mathrm{det}{\tilde S} = - r^2 (Kr)^2 = 0\,,
\ee
which proves the existence of a goldstone mode in a basis-invariant fashion.

\section{Discussion and conclusions}

The principal result of this paper is a demonstration that the mass spectrum of the general 2HDM
can be studied in a reparametrization-invariant way within the Minkowski-space formalism
of \cite{ivanov1,ivanov2}. 
This means that the scalar propagators can be now written explicitly and can be used, for example,
to improve the thermal one-loop calculations of \cite{thermal-general}.

Another interesting issue that one can now address is to understand to what extent the perturbativity/tree-level unitarity
bounds on the Higgs potential restrict the values of the Higgs boson masses.
In the Standard Model, there is a strong correlation between the value of the quartic coupling constant $\lambda$
and the Higgs boson mass. Therefore, an upper limit on $\lambda$ implies a corresponding upper limit on $M_H$.
In the 2HDM, due to a large number of free parameters, the situation is more complicated, see 
\cite{akeroyd,ginzreview,kanemura,GIunitarity,kladiva}. 
It was noted that in certain cases masses of some of the Higgs bosons can be very high without violating 
the tree-level unitarity conditions. With an explicit expression for the trace of the mass matrix,
one could now attack this problem in the most general case within the Minkowski-space technique.
The only piece still missing is a reparametrization-covariant expression for the tree-level unitarity constraints. 

In conclusion, we showed that the Minkowski-space approach to the most general 2HDM can also be used
to analyze the mass spectrum of the physical Higgs bosons.
We calculated the traces of the powers of the mass matrix and its determinant for all types of 
vacuum that can exist in 2HDM.
These results can now be used to get even more insight into the properties of the general 2HDM.
\\

We are thankful to J.-R. Cudell for helpful discussions. 
This work was supported by the Belgian Fund F.R.S.-FNRS via the
contracts of Charg\'e de recherches (I.P.I.) and of Aspirant (A.D.).
The work of I.P.I. was in part supported by grants
RFBR 08-02-00334-a and NSh-1027.2008.2

\appendix

\section{Algebra of matrices $\Sigma^\mu$ and $\Pi^\mu$}

The four-vector of matrices $\Sigma^\mu$ is introduced via Eq.~(\ref{definitionSigma}).
The full 8-by-8 matrices $\Sigma^\mu$ have block-diagonal form and are built from two identical 4-by-4 
matrices, which we also denote by the same letter $\Sigma$'s and whose properties we describe here. 

$\Sigma^0$ is just the unit matrix, while the explicit expressions of $\Sigma^i$ are:
\be
\Sigma^1 = 
\left(\begin{array}{cccc} 
0 & 0 & 1 & 0\\
0 & 0 & 0 & 1\\
1 & 0 & 0 & 0\\
0 & 1 & 0 & 0
\end{array}\right)\,,\quad
\Sigma^2 = 
\left(\begin{array}{cccc} 
0 & 0 & 0 & 1\\
0 & 0 & -1 & 0\\
0 & -1 & 0 & 0\\
1 & 0 & 0 & 0
\end{array}\right)\,,\quad
\Sigma^3 = 
\left(\begin{array}{cccc} 
1 & 0 & 0 & 0\\
0 & 1 & 0 & 0\\
0 & 0 & -1 & 0\\
0 & 0 & 0 & -1
\end{array}\right)\,.\label{explicitSigma}
\ee
These matrices satisfy the Clifford algebra condition:
\be
\{\Sigma^i,\Sigma^j\} = 2 \delta^{ij}\mathbb{I}_4\,.
\ee
The set of $\Sigma$'s is not closed under taking commutators. Instead, they can be expressed via 
real antisymmetric matrices $\Pi^i$:
\be
\Pi^i \equiv \Pi^0\Sigma^i\,, \quad\mbox{where}\quad
\Pi^0 = \left(\begin{array}{cccc} 
0 & 1 & 0 & 0\\
-1 & 0 & 0 & 0\\
0 & 0 & 0 & 1\\
0 & 0 & -1 & 0
\end{array}\right)\,.
\ee
The matrix $\Pi^0$ is the generator of the simultaneous $SO(2)$ rotations between the real and imaginary parts the two doublets;
it commutes with all $\Sigma^i$ and its square is equal to $-1$.
The set of matrices $\Sigma^i$ and $\Pi^i$ now forms the algebra:
\be
[\Sigma^i,\Sigma^j] = 2\epsilon^{ijk}\Pi^k\,,\quad 
[\Sigma^i,\Pi^j] = -2\epsilon^{ijk}\Sigma^k\,,\quad
[\Pi^i,\Pi^j] = -2\epsilon^{ijk}\Pi^k\,.
\ee
Note that $\Pi^i$ do form a closed algebra.

The algebra of $\Sigma^i$ and $\Pi^i$ is isomorphic to
the usual Poincar\'e algebra of the generators of boosts and rotations. 
Using this, we can introduce matrices 
\be
X_\pm^i = {1 \over 4}\left(\pm \Sigma^i - i\Pi^i\right)\,,
\ee
which satisfy the following commutation laws:
\be
[X_\pm^i,X_\pm^j] = i \epsilon^{ijk}X_\pm^k\,,\quad
[X_\pm^i,X_\mp^j] = 0\,.
\ee 
Finally, we note that any four-vector $a_\mu$ can be associated with a real symmetric matrix $A = a_\mu \Sigma^\mu$, which has the following properties:
\be
\mathrm{det}A = (a_\mu a^\mu)^2\,,\quad A^{-1} = {a_\mu \bar\Sigma^\mu \over a_\mu a^\mu}\,,\quad
\mbox{with} \quad \bar\Sigma^\mu \equiv (\Sigma^0,\,-\Sigma^i)\,.
\ee

\end{document}